\documentclass[aps,prb,reprint,showpacs,superscriptaddress,longbibliography]{revtex4-2}
\usepackage{mathtools,amssymb,graphicx,units}

\usepackage[plainpages=false,pdfpagelabels,colorlinks=true,linkcolor=red,urlcolor=blue,citecolor=blue,pdftitle={Title},pdfauthor={Rubem Mondaini},pdfdisplaydoctitle=true,pdfduplex=DuplexFlipLongEdge]{hyperref}
\usepackage{amsmath,graphics,units}
\usepackage[table,xcdraw]{xcolor}
\usepackage{verbatim}
\usepackage[english]{babel}
\usepackage{color}
\usepackage{ulem}
\usepackage{physics}

\begin{document}
\title{Non-Hermitian Haldane-Hubbard model: Effective description of an open system with balanced gain and loss}
\author{Tian-Cheng Yi}
\affiliation{Department of Physics,
Zhejiang Sci-Tech University, Hangzhou 310018, China}

\affiliation{Zhejiang Key Laboratory of Quantum State Control and Optical Field Manipulation,
Zhejiang Sci-Tech University, Hangzhou 310018, China}

\affiliation{Beijing Computational Science Research Center, Beijing 100193, China}
\author{Rubem Mondaini}
\email{rmondaini@uh.edu}
\affiliation{Department of Physics, University of Houston, Houston, Texas 77004, USA}
\affiliation{Texas Center for Superconductivity, University of Houston, Houston, Texas 77204, USA}

\begin{abstract}
We study the correlated Haldane-Hubbard model with single-particle gain and loss, focusing on its non-Hermitian phase diagram and the ensuing non-unitary dynamic properties. The interplay of interactions and non-hermiticity results in insulating behavior with a phase diagram divided into three distinct regions, exhibiting either topologically gapped or (real) gapless regimes and a trivial phase. The latter is mapped by the emergence of a local order parameter associated with a charge density wave. A ${\cal PT}$-symmetry breaking at the low-lying spectrum occurs when increasing the gain-loss magnitude at a fixed interaction strength, marking the transition from gapped to gapless topological behavior. Further increase leads to the onset of charge ordering in a first-order phase transition in which level crossing takes place in the spectrum's imaginary part. The support that the staggered gain and loss display to robust charge density wave in equilibrium is confirmed in the real-time dynamics in the presence of non-hermiticity, suggesting that engineered gain and loss can be used to tailor an ordered many-body state in experiments.
\end{abstract}

\maketitle
\section{Introduction}
Non-hermiticity naturally arises in the Hamiltonians of systems coupled to an environment where energy, particle, or information exchange takes place~\cite{El-Ganainy2018, Ashida2020, Bergholtz2021}. Since the assumption of truly isolated systems -- often made in theoretical models -- rarely holds in realistic physical scenarios, the study of non-Hermitian physics becomes essential. In particular, the coupling to external degrees of freedom can be experimentally tuned in a variety of settings~\cite{Feng2014, Wu2019, Zhao2019, Helbig2020, Ozturk2021, Ding2021, Xu2023}, enabling precise investigations of dissipation effects, including those in strongly correlated systems. Simultaneously, the emergence of novel phenomena under these conditions, such as asymmetric edge modes, dubbed the non-Hermitian skin effect~\cite{Xiao2020}, has motivated the generalization of topological state classification~\cite{Zhou2019, Kawabata2019, Kawabata2019b} and their corresponding invariants to the non-Hermitian regime, via either modifying the domain of definition of topological invariants from the Brillouin zone to the generalized Brillouin zone~\cite{Yao2018, Yokomizo2019} or through continuous transformations of the complex
energy spectrum~\cite{Gong2018}.

A particular example of the investigation of such a non-hermiticity and topology interplay was done for the paradigmatic Haldane model~\cite{Haldane1988} in the case that the coupling to an environment encompasses both gain and losses~\cite{Resendiz-Vasquez2020, Li2022, Sarkar2023}. Its original Hermitian form is characterized by the formation of protected edge modes, whose number and chirality are mapped by a $\mathbb{Z}_2$ topological invariant, the Chern number. The introduction of balanced gain-loss non-Hermiticity leads to higher-order skin modes, i.e., reduced skin dimensionality modes, in this case, associated with corner localization~\cite{Li2022}, a state that can emerge even in the case of non-hermiticity derived from non-reciprocity~\cite{Lee2019, Zou2021}. This leads us to a key open question: To what extent are these non-Hermitian topological phenomena robust in the presence of strong interactions?

Before answering that, let us review some known results of the interacting Haldane model in Hermitian settings. The Haldane model has been extended to explore the effects of different types of interactions in either spinless~\cite{Varney2010, Varney2011} and spinful settings~\cite{Imriska2016, Zheng2015, Vanhala2016, Tupitsyn2019, Shao2021}, which can lead to spontaneous breaking of SU(2) symmetry \cite{Vanhala2016,Tupitsyn2019,Yuan2023,He2024}, or even the development of topological properties with increased interactions at certain electron fillings~\cite{Mai2023}. In this context of correlated systems, introducing disorder might result in a phase of topological Anderson insulators~\cite{Yi2021}, while hopping dimerization can give rise to higher-order topological phases~\cite{Yi2023}. Additionally, methods have been devised to manipulate real-time dynamics through carefully tuned time-dependent perturbations to reach target states with non-trivial topology~\cite{Shao2021b}.

The extension to non-hermiticity of the Haldane-Hubbard model was only once investigated by us, where we unveiled the effects of dissipation only for either one or two-particle channels~\cite{Wang2023}, inspired by natural effects expected in the emulation of the Haldane model in trapped ultracold atom settings~\cite{Jotzu2014}. In this case, both dissipation types result in the exponential decay of particle density over time, suppressing local charge ordering, typically induced by the interactions.  On the other hand, incorporating non-hermiticity via gain and loss is crucial for uncovering novel quantum phenomena and informing advanced technological applications. Here, the focus will thus be on a regime with balanced gain and loss, such that, on average, the density remains constant.

The presentation structure is as follows: In Sec.~\ref{sec:model}, we introduce the gain-loss Haldane-Hubbard model and the corresponding physical quantities.  Section~\ref{sec:results} establishes the ground-state phase diagram in equilibrium. We also analyze the low-energy structure of the model in this context, specifying the nature of the transitions taking place. In Sec.~\ref{sec:dynamics}, we investigate the dynamics properties of the model, that is, in the presence of a bath that induces simultaneous gain and loss, and focus on the fate of the charge ordering. Finally, Sec.~\ref{sec:summary} summarizes our findings.

\section{Model and quantities} 
\label{sec:model}
\begin{figure*}[htp]\centering
\includegraphics[width=2\columnwidth]{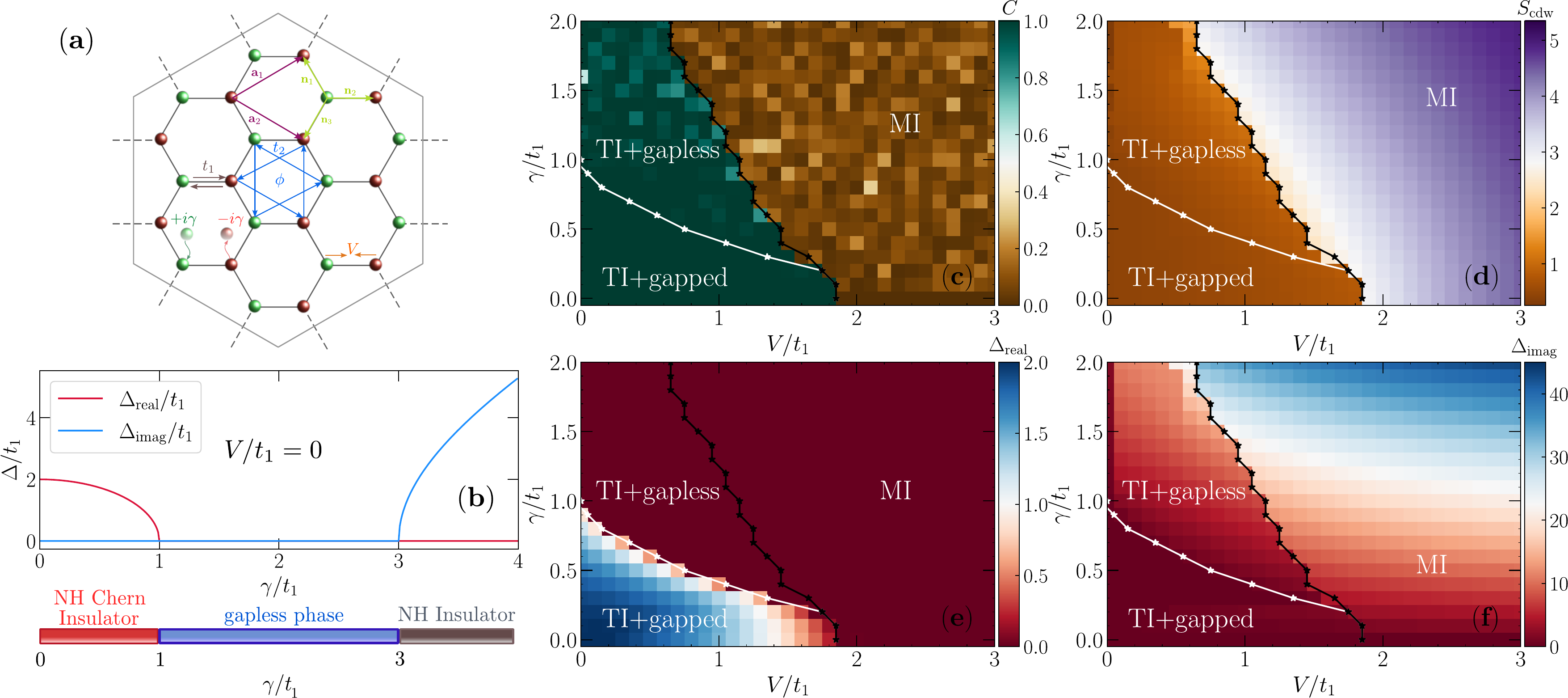}
\caption{
(a) Schematic diagram of the non-Hermitian Haldane-Hubbard model with its relevant terms annotated in the six-fold rotationally symmetric lattice with $N_s=24$ sites (primitive and nearest-neighbor vectors are also marked); the dissipation and gain rates are given both by $\gamma$.
(b) The evolution of the real ($\Delta_{\rm real}$) and imaginary parts ($\Delta_{\rm imag}$) of the excitation gap in the non-interacting limit, $V/t_1=0$, with the gain-loss amplitude $\gamma$. The bottom cartoon identifies the two transitions at $\gamma/t =1$ and 3, from a non-Hermitian topological insulator to a gapless regime and further to a topologically trivial non-Hermitian insulator, respectively.
(c)--(f) The phase diagram of the model, classified by the computation of four physical quantities governed by the gain-loss parameter $\gamma$ and the interaction strength $V$ --- different quantities are mapped by the corresponding color bars. (c) The Chern number $C$; (d) the charge density wave structure factor $S_{\rm cdw}$; (e) the real part of the energy gap $\Delta_{\rm real}$ and (f) the corresponding imaginary part of the energy gap $\Delta_{\rm imag}$. The distinct phases are the topological insulator phases (TI + gapped and TI + gapless) and the Mott insulator phase (MI). The transition between these phases is governed by the gain-loss parameter $\gamma$ and the interaction strength $V$. The connected star markers identify an estimated transition location among the three phases based on compiling the four quantities for the 24-site cluster.
}
\label{fig:Fig_1}
\end{figure*}

We study the spinless Haldane-Hubbard in the presence of gain and loss, whose Hamiltonian reads,
\begin{align}
    \hat H = &-t_1\!\!\sum_{\langle l,m\rangle}\!\!\left(\hat c_l^\dagger \hat c_m^{\phantom{\dagger}} + {\rm H.c.}\right)-t_2\!\!\sum_{\langle\langle l,m\rangle\rangle}\!\!\left(e^{i\phi_{lm}}\hat c_l^\dagger \hat c_m^{\phantom{\dagger}} + {\rm H.c.}\right) \nonumber \\
    &+ V\sum_{\langle l,m\rangle}\hat n_l\hat n_m + i\gamma\sum_{l}(-1)^l \hat n_l\ .
    \label{eq:H_haldane_hubbard}
\end{align}
Here, $\hat c_l^\dagger$ ($\hat c_l^{\phantom{\dagger}}$) is the fermionic annihilation (creation) operator at orbital $l$ of a honeycomb lattice displaying $N_s$ sites; $\hat n_l = \hat c_l^\dagger \hat c_l^{\phantom{\dagger}}$ is the number operator. Nearest neighbor, $\langle l,m \rangle$, and next-nearest neighbor, $\langle\langle l,m\rangle\rangle$, hoppings with magnitudes $t_1$ and $t_2$, respectively, promote itinerancy in the lattice. The latter picks up a phase $e^{i\phi_{lm}}$ with $\phi_{lm} = +\phi$($-\phi$) for counter-clockwise (clockwise) hoppings [see Fig.~\ref{fig:Fig_1}(a)]. The nearest-neighbor interaction strength is mapped by $V$, and the non-hermiticity is introduced by a purely imaginary staggered potential of amplitude $\gamma$, emulating an engineered gain and loss process.

We quantify the low-lying spectral properties of \eqref{eq:H_haldane_hubbard} using a Krylov-Schur exact diagonalization method~\cite{Petsc, Slepc}, assigning the ground-state as the state with the smallest real part. We further focus on the right eigenvectors $\hat H |\alpha^R\rangle = E_\alpha|\alpha^R\rangle$, as opposed to the left ones, $\hat H^\dagger |\alpha^L\rangle = E_\alpha^*|\alpha^L\rangle$, since topological invariants are the same irrespective of the choice~\cite{Shen2018, Xu2020} and, additionally, local quantities exhibit qualitatively similar results~\cite{Wang2023}.

To quantify the topological nature of the ground state in exact diagonalization (ED), we compute the Chern number $C$ by using a discretized form of the Berry curvature integration~\cite{Fukui05, Varney2010, Varney2011}. This is given by:
\begin{equation}
    C = \int \frac{d\phi_x d\phi_y}{2 \pi {i}} \left( \langle\partial_{\phi_x}
      \Psi^\ast | \partial_{\phi_y} \Psi\rangle - \langle{\partial_{\phi_y}
      \Psi^\ast | \partial_{\phi_x} \Psi\rangle} \right),
      \label{eq:Chern}
\end{equation}
where twisted boundary conditions $\{\phi_x, \phi_y\}$ are applied~\cite{Poilblanc1991, Niu1985}.

In the Hermitian regime ($\gamma\to 0$), a topological phase transition occurs at large repulsive interaction strengths $V$, transforming the topological insulator into a trivial charge-density-wave (CDW) insulator~\cite{Varney2010, Varney2011}. In the limit $V \to\infty$, the ground state is a perfect CDW, where one of the two sublattices is occupied while the other is empty, leaving lattice translational symmetry intact but breaking inversion symmetry~\cite{Wessel2007}. To characterize it, we compute the ${\bf k}=0$ CDW structure factor,
\begin{align}
  \label{eq:struct}
  S_{\rm cdw}  &\equiv \frac{1}{N_s}\sum_{l,m} c({\bf r}_l - {\bf r}_m),
\end{align}
with density-density correlations
\begin{align}
  c({\bf r}_l - {\bf r}_m) &= \langle (\hat n_l^A - \hat n_l^B) (\hat n_m^A - \hat n_m^B) \rangle,
\end{align}
where $\hat n_l^A$ and $\hat n_l^B$ are the number operators on sublattices $A$ and $B$
in the $l$-th unit cell of the honeycomb lattice, respectively. 

Immediate verification of a first-order phase transition, invariably tied to the modification of the topological invariant, is obtained by the computation of the excitation (or many-body) gap:
\begin{eqnarray}
    \Delta = E_1(N_s/2)- E_0(N_s/2),
\end{eqnarray}
which quantifies the energy difference between the two-lowest eigenvalues of Eq.~\ref{eq:H_haldane_hubbard} for the studied filling factor $N_e = N_s/2$ (half-filling). Since for $\gamma\neq 0$ the spectrum is not guaranteed to be real, this gap thus acquires real and imaginary parts. In what follows, we set $t_1$ as our unit of energy, with $t_2 = 0.2t_1$ and $\phi=\pi/2$. The first choice makes it easier to compare with previous literature that establishes that in the Hermitian case ($\gamma=0$), the first-order topological-to-trivial phase transition with the accompanying onset of CDW order takes place at a critical interaction $V_c/t_1 \simeq 1.9$~\cite{Varney2010, Varney2011}. The second maximizes the robustness of the topological insulating phase upon the perturbation from other parameters.

\section{Results}
\label{sec:results}

We start by revising the effects of non-hermiticity in the non-interacting limit ($V=0$) --- previous research has focused on this regime~\cite{Resendiz-Vasquez2020, Li2022}. In such case, Eq.~\eqref{eq:H_haldane_hubbard} can be diagonalized in momentum space $\hat H(V=0) \equiv \hat H_0 = \sum_{\bf k}\psi_{\bf k}^\dagger\hat H_{\bf k}^{\phantom{\dagger}}\psi_{\bf k}^{\phantom{\dagger}}$, resulting in two bands (see details in the Appendix), $E_{{\bf k}}^\pm \!=\! A({\bf k}) \!\pm\! \sqrt{|f({\bf k})|^2\!+\!\Delta m({\bf k})^2 \!-\! \gamma^2 \!+\! 2i\gamma \Delta m({\bf k})}$, exhibiting finite imaginary components; $A$, $f$ and $\Delta m$ are real functions of their arguments. As a result, the gap at half-filling, $\Delta \equiv E_{{\bf k}}^+ - E_{{\bf k}}^-=\Delta_{\rm real} + i\Delta_{\rm imag}$, acquires both real and imaginary parts as a function of the gain-loss parameter $\gamma$. Figure~\ref{fig:Fig_1}(b) compiles this result, depicting that at $\gamma/t_1 = 1$ the real part of the gap $\Delta_{\rm real}$ closes whereas at $\gamma/t_1 = 3$ the imaginary, $\Delta_{\rm imag}$, opens. This establishes three phases in the non-interacting phase diagram: (i) $0\leq \gamma/t_1 < 1$, a non-Hermitian Chern insulator, with $\Delta_{\rm real}>0$ and $\Delta_{\rm imag}=0$; (ii) $0\leq \gamma/t_1 < 1$ a gapless phase in which $\Delta_{\rm real}=\Delta_{\rm imag} =0$ and (iii) $\gamma/t_1 \geq 3$ a trivial non-Hermitian insulator with $\Delta_{\rm real}=0$ and $\Delta_{\rm imag} >0$. Notably, the non-Hermitian Chern insulator is adiabatically connected to its Hermitian counterpart in the $\gamma \to 0$-limit~\cite{Haldane1988}, exhibiting a quantized Chern number even in the presence of non-hermiticity~\cite{Li2022}.

\begin{figure}[t!]\centering
\includegraphics[width=1\columnwidth]{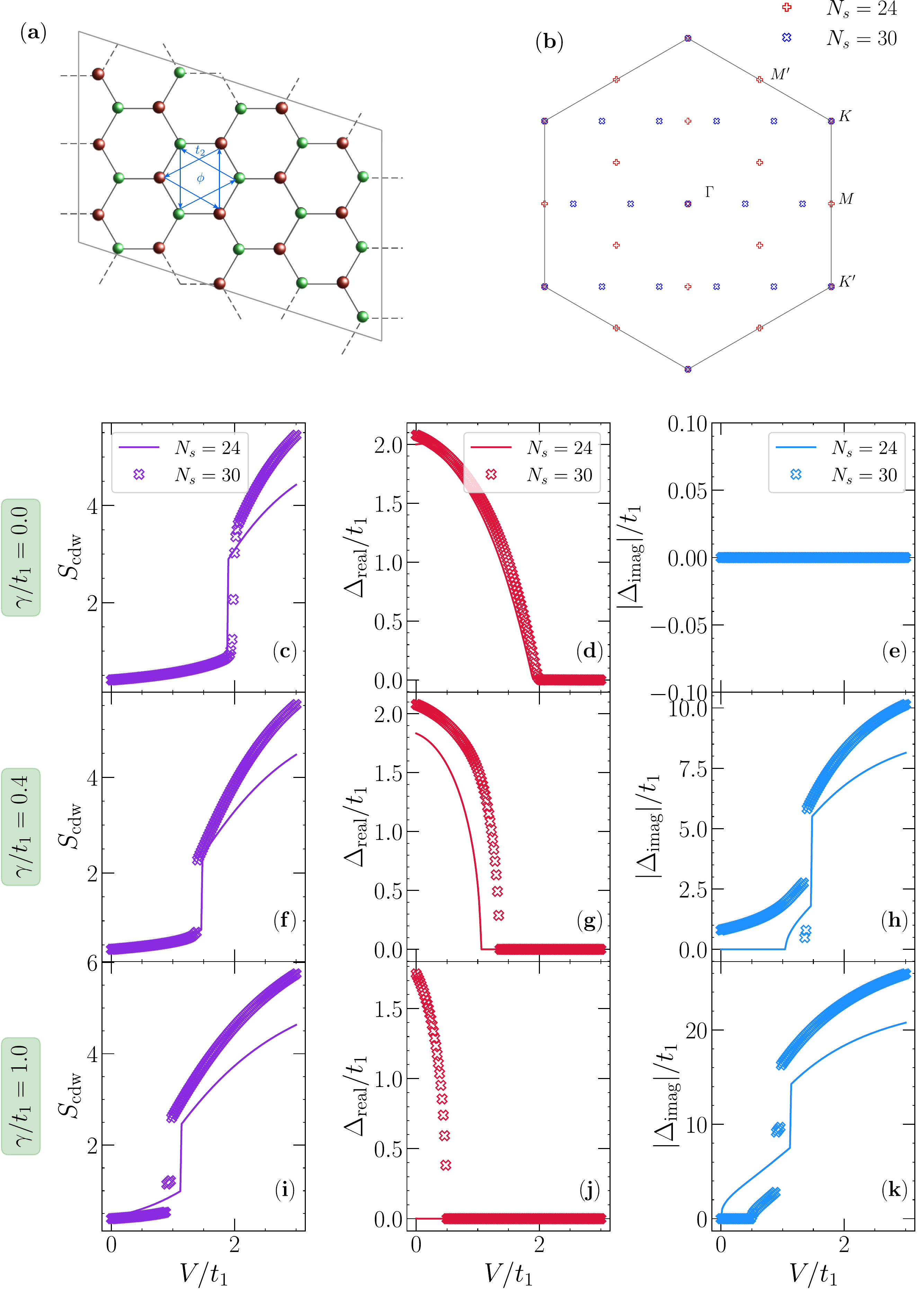}
\caption{(a) Cartoon of the $N_s=30$ cluster used to study finite-size effects. (b) A comparison of the allowed momenta in the two clusters investigated with high-symmetry points annotated in the first Brillouin zone -- notice that some of these points are equivalent by a reciprocal lattice translation but are repeated for clarity. (c), (f) and (i) show the interaction strength dependence of the CDW structure factor for gain-loss ratios $\gamma/t_1 = 0, 0.4$ and 1.0, respectively. Panels [(d), (g), (j)] and [(e), (h), (k)] show the same for the real and imaginary parts of the excitation gap, $\Delta_{\rm real}$ and $\Delta_{\rm imag}$, respectively. Continuous lines (markers) refer to the $N_s=24$ ($N_s=30$) cluster.}
\label{fig:Fig_3}
\end{figure}

For finite interaction strengths, one obtains the ground-state properties numerically -- the phase diagram is mapped in Figs.~\ref{fig:Fig_1}(c--f) for a lattice size featuring $N_s=24$, highlighted in Fig.~\ref{fig:Fig_1}(a). Such a lattice exhibits all the point-group symmetries of the one in the thermodynamic limit, leading to reduced overall finite-size effects; results will be contrasted later with a different system size in Fig.~\ref{fig:Fig_3}.

\begin{table}[b!]
\caption{
The characteristics of the three phases for finite interactions. Here, C.P. denotes a complex conjugate pair.}
\label{table:phase_characteristics}
\begin{ruledtabular}
\begin{tabular}{ c c c c c c}
phase name	& $C$ & CDW order & $\Delta_{\rm real}$ & $\Delta_{\rm imag}$ & $E_0$\\
\hline
TI (gapped) &$1$ & no & $>0$ & $=0$ & real\\
TI(real gapless) &$1$ & no & $=0$ & $>0$ & C.P.\\
MI &$0$ & yes & $=0$ & $>0$ & C.P.
\end{tabular}
\end{ruledtabular}
\end{table}

As depicted in Fig.~\ref{fig:Fig_1}, the phase diagram of the model is systematically divided into three regions, each identified by a distinct combination of the physical quantities within the $(\gamma, V)$-parameter space -- a summary is given in Table~\ref{table:phase_characteristics}. The distribution of the topological invariant $C$ [Fig.~\ref{fig:Fig_1}(a)] and the charge density wave structure factor $S_{\rm cdw}$ [Fig.~\ref{fig:Fig_1}(b)] emphasize that topology and charge ordering~\footnote{We highlight that long-range order can only be obtained by showing that $S_{\rm cdw}$ is extensive in the system size while approaching the thermodynamic limit; here this analysis is taken thus as merely qualitative.} are mutually exclusive, extending a result that originates in the Hermitian regime~\cite{Varney2010, Varney2011} to the presence of gain and loss. Such an incompatibility of local and non-local orders also occurs when only losses are considered in a similar Hamiltonian~\cite{Wang2023}. Additionally, resolving the energy gap $\Delta$ into its real and imaginary parts, $\Delta_{\rm real}$ [Fig.~\ref{fig:Fig_1}(e)] and $\Delta_{\rm imag}$ [Fig.~\ref{fig:Fig_1}(f)], helps in discerning two different phases within the topological interacting region. While the topological insulating regime at small gain-loss strengths exhibits a gapped real spectrum with gapless imaginary parts, a direct extension of its non-interacting limit, increasing $\gamma$ leads to $\Delta_{\rm real}\to 0$ while at the same time $\Delta_{\rm imag}$ turns finite. Thus, such a `partially' gapless regime finds no parallel in the $V=0$ limit, featuring $\Delta_{\rm real}=\Delta_{\rm imag}=0$ at that regime of parameters, yet being still topological.

\begin{figure}[t!]\centering
\includegraphics[width=1\columnwidth]{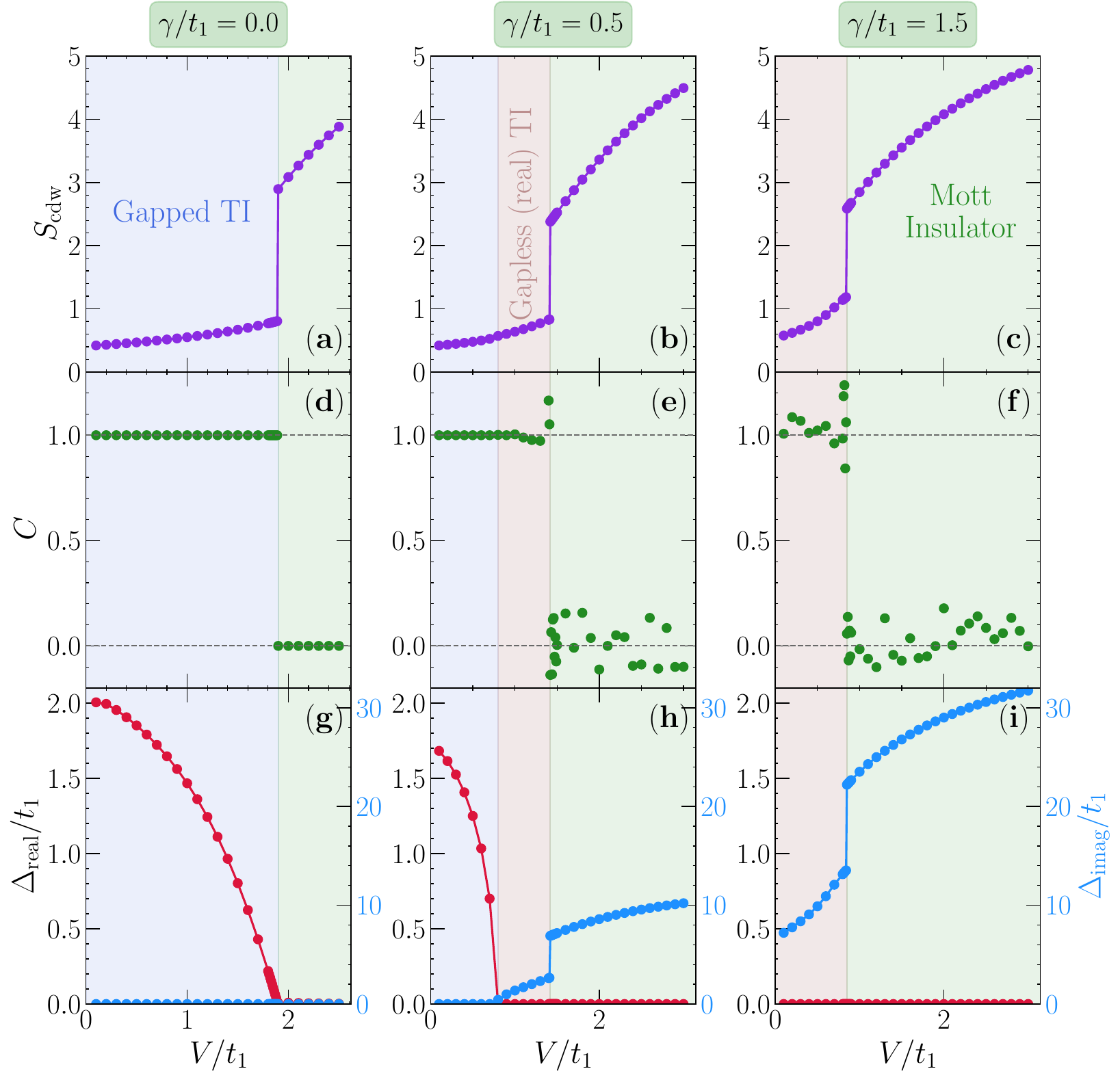}
\caption{
Line cuts of the phase diagram of the non-Hermitian Haldane-Hubbard at different gain-loss ratios, $\gamma/t_1 = 0, 0.5$, and 1.5. 
The three rows show different physical quantities across the interaction strength $V$: (a--c) charge density wave structure factor $S_{\rm cdw}$, (d--f) topological Chern number $C$, and (g--i) real and imaginary parts of the energy gap, $\Delta_{\rm real}$ and $\Delta_{\rm imag}$; if $\Delta_{\rm real}=0$, $\{E_n\}$'s are ordered according to their imaginary part such that $\Delta_{\rm imag}>0$. Each column corresponds to a different value of $\gamma$.} 
\label{fig:Fig_2}
\end{figure} 

A close inspection of the different quantities at specific values of the gain-loss parameter $\gamma$ for increasing interactions $V$ helps characterize the various phases, and further delineates their boundaries. Figure~\ref{fig:Fig_2} thus presents line cuts of the phase diagram at representative values of the gain-loss parameter at $\gamma/t_1 = 0$, $0.5$, and $1.5$. As previously investigated, the Hermitian limit ($\gamma = 0$) is identified by a single phase transition with the interaction strength at a critical value $V_c \simeq 1.9t_1$~\cite{Varney2010, Varney2011}. Such a concomitant topological-to-trivial and quantum disordered-to-CDW phase transition is first-order: The excitation gap (here only real due to hermiticity) closes, and physical quantities, such as $S_{\rm cdw}$ display discontinuous behavior from a small to a larger value.

With finite $\gamma$, on the other hand, this single transition splits in two: a first one at smaller interactions at which $\Delta_{\rm real}$ vanishes~\footnote{We note that in the non-Hermitian case, $\gamma\neq 0$, $\Delta_{\rm real} = 0$ in the trivial insulating phase even in a finite cluster, since the ground-state forms a complex-conjugate pair with another state in this phase. For $\gamma=0$, this is never the case in a finite cluster; the two low-lying energy eigenstates feature CDW ordering in this phase but are even and odd under parity symmetry. Degeneracy only occurs in the thermodynamic limit, leading thus to $\Delta_{\rm real}\to 0$.} and a second one at larger $V$'s where $S_{\rm cdw}$ exhibits the characteristic discontinuity associated with the onset of CDW order. Meanwhile, the first transition marks the point at which the imaginary gap opens while at the second, $\Delta_{\rm imag}$ exhibits a discontinuity characteristic of first-order phase transitions, but here, as displayed in the imaginary part of the gap. The corresponding Chern number $C$, Figs.~\ref{fig:Fig_2}(d), \ref{fig:Fig_2}(e) and \ref{fig:Fig_2}(f), makes it clear that this second transition is from a (non-Hermitian) topological regime to a trivial one -- fluctuations from perfect quantization can emerge because the necessary gap opening condition in the torus of twisted boundary conditions used to compute it is not always satisfied~\cite{Fukui05}, even more so because $\Delta_{\rm real} = 0$ in the topologically trivial phase.

Next, we investigate the influence of finite-size effects in these results. For that, we study a cluster featuring $N_s=30$ [Fig.~\ref{fig:Fig_3}(a)], which is also bipartite considering the periodic boundary conditions. We start by noticing that the main characteristics of CDW structure factors are preserved and that in the large interaction strength, $S_{\rm cdw}$ grows with the system size, indicative of CDW long-range ordering. Despite minor quantitative differences in the critical interaction strengths that trigger this ordering, the results are the same qualitatively, see Figs.~\ref{fig:Fig_3}(c), \ref{fig:Fig_3}(f) and \ref{fig:Fig_3}(i).

\begin{figure}[t!]\centering
\includegraphics[width=0.9\columnwidth]{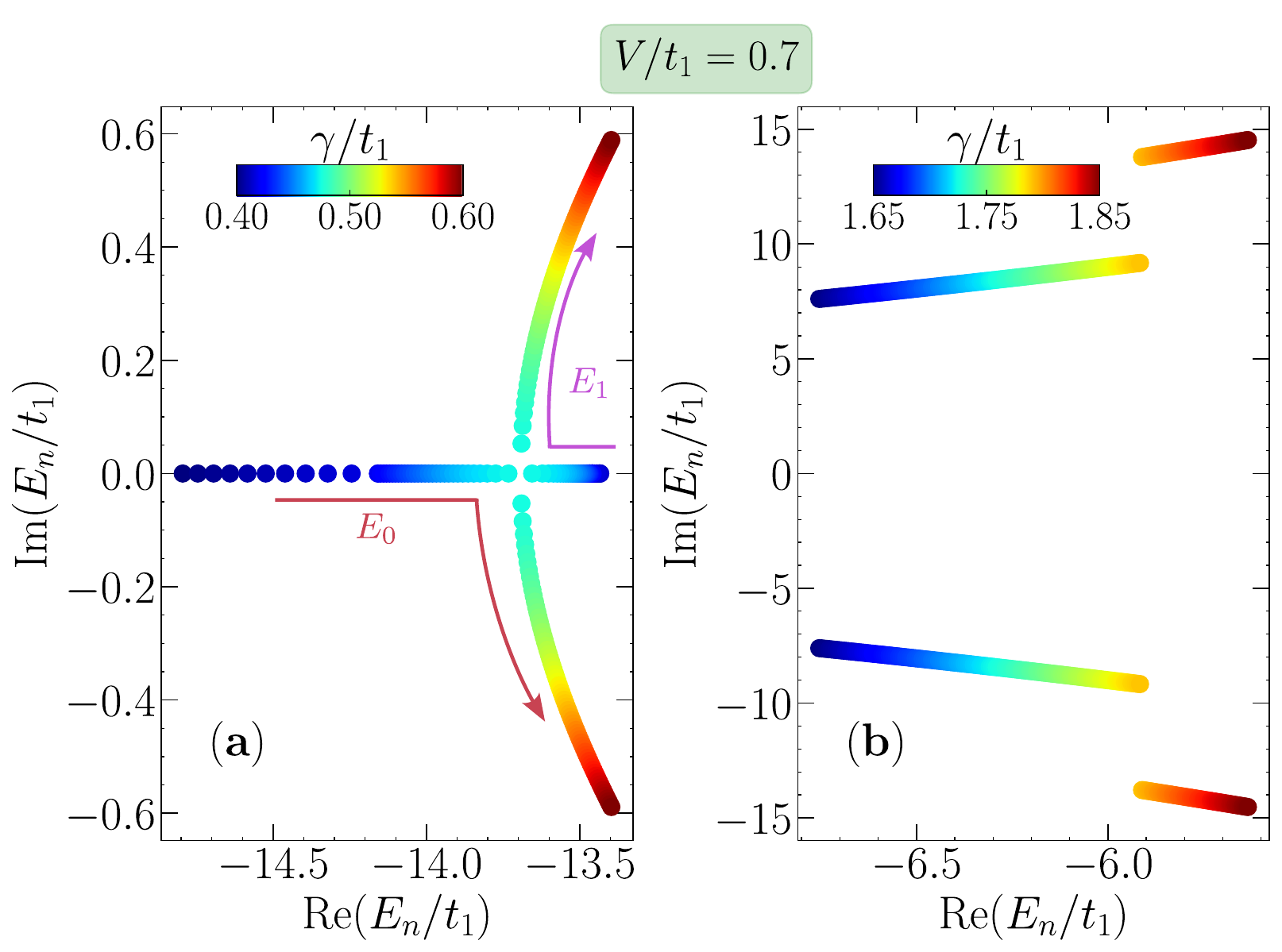}
\caption{Energy spectrum analysis of the two states with the smallest ${\rm Re}(E_n)$ as a function of the gain-loss parameter $\gamma$, mapped by the color bar. In (a), we focus on the weak non-hermiticity regime, where a transition occurs between the gapped TI and the `partially'-gapless TI. The same is highlighted for the topological-to-trivial transition at stronger non-hermiticity in (b). Here, the interaction strength is set at $V/t_1=0.7$.}
\label{fig:Fig_4}
\end{figure}

The same, however, cannot be concluded about the real and imaginary gaps at finite $\gamma$. Focusing on the large non-hermiticity regime, $\gamma=t_1$, we notice that while $\Delta_{\rm real}$ is finite for the $N_s =30$ cluster when $V\to 0$ [Fig.~\ref{fig:Fig_4}(j)], such a gap is absent in the $N_s = 24$ lattice results. To explain this fundamental discrepancy, we note that at the non-interacting regime, $\gamma/t_1=1$ is the point that separates the real-gapped TI and gapless TI regimes [see Fig.~\ref{fig:Fig_1}(b)]. According to the analysis in the Appendix, this is mapped by a gap closing at the high-symmetry point $M$ of the first Brillouin zone. Since this is not a valid momentum point for the $N_s =30$ cluster, unlike for $N_s=24$ [see Fig.~\ref{fig:Fig_3}(b)], the results exhibit \textit{systematic} finite-size effects in this case. This corroborates the argument that finite-size effects at small clusters can be rather sensitive to the associated symmetries of the clusters used~\cite{Varney2010}. Nonetheless, it does not substantially affect the onset of the charge-ordered regime because, in this transition, the most relevant momenta are $K$ and $K^\prime$, wherein the gap opening occurs. The two clusters we investigate exhibit these high-symmetry points [Fig.~\ref{fig:Fig_3}(b)], and as a result, the finite-size effects are small for $S_{\rm cdw}$.

Having shown the characteristics of the different regimes in the phase diagram, we can now closely analyze the interacting phase transitions with growing gain-loss ratio $\gamma$ by inspecting the corresponding complex low-lying spectrum. In particular, this analysis allows one to identify parity-time (${\cal PT}$) symmetry breaking, associated with the regime at which two eigenmodes coalesce in the complex plane, defining an exceptional point~\cite{Ashida2020}. Figure~\ref{fig:Fig_4} focuses on the ground-state $E_0$ and the first excited state $E_1$, that is, the two eigenstates in the energy spectrum with the lowest real part of their energies. The first transition, assigned as a point in which the $\Delta_{\rm real}$ closes (and $\Delta_{\rm imag}$ opens), is thus clearly one where a ${\cal PT}$-symmetry breaking takes place: both $E_0$ and $E_1$ are identically real, and with increasing $\gamma$, it leads to an exceptional point [at around $\gamma_{\rm ep} = 0.53t_1$ for $V=0.7t_1$, Figure~\ref{fig:Fig_4}(a)] where these eigenvalues coalesce [see Fig.~\ref{fig:Fig_4}(a)]~\footnote{We note that such a ${\cal PT}$-symmetry breaking associated with just two-eigenvalues in the low-lying spectrum was also observed in other interacting non-Hermitian models, such as the one-dimensional interacting Hatano-Nelson model~\cite{Zhang2022}}. Further enhancement of the gain-loss parameter $\gamma$ makes $E_0$ and $E_1$ become complex-conjugate pairs. Consequently, $\Delta_{\rm real} = 0$ and $\Delta_{\rm imag} = 2 |{\rm Im}(E_{0,1})|$, thus identifying the (real) gapless TI phase we previously described.

Finally, at even larger values of $\gamma$ for a fixed interaction strength, the non-Hermitian MI phase emerges [at $\gamma = 1.79t_1$ for $V=0.7t_1$, Figure~\ref{fig:Fig_4}(b)]. Here, the low-lying spectrum is identified by a typical first-order phase transition at which the real part of the eigenvalues exhibits a level crossing, and, generalized by the non-hermiticity, the corresponding imaginary parts display a sudden jump, thus explaining the discontinuity in the results for $\Delta_{\rm imag}$ at large $\gamma's$ in Figs.~\ref{fig:Fig_2} and \ref{fig:Fig_3}.

\begin{figure*}[t!]\centering
\includegraphics[width=1\textwidth]{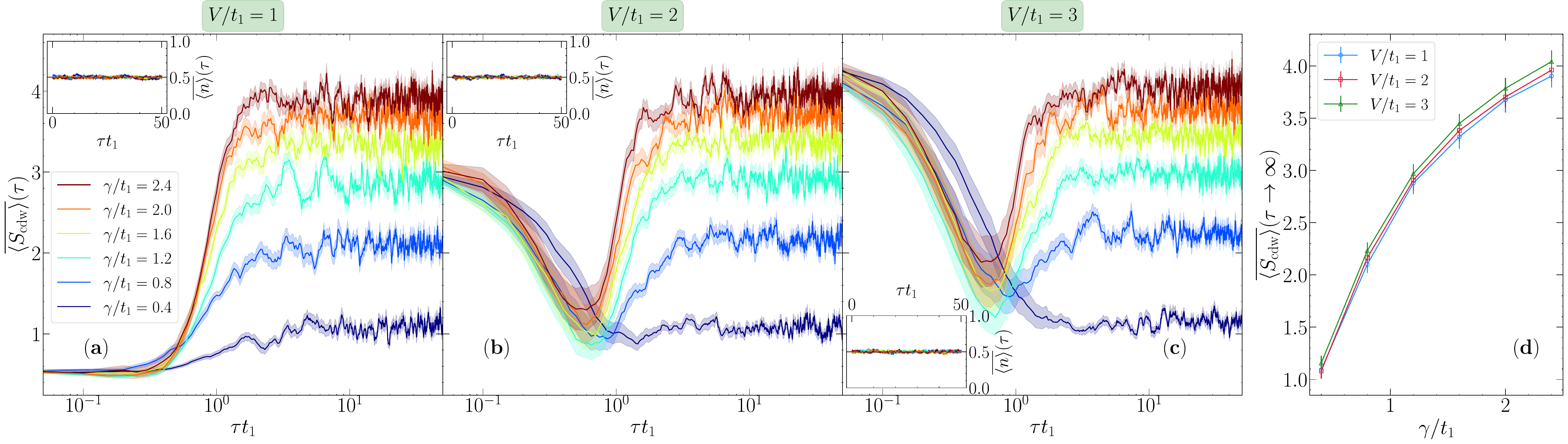}
\caption{Dynamics of the CDW structure factor with growing gain-loss strength and interactions (a) $V/t_1=1$, (b) $V/t_2=2$, and (c) $V/t_1=3$ while averaging 50 trajectories in the dynamics stochastic process; the insets show the corresponding densities, displaying minor oscillations about its initial value $\langle\hat n\rangle = 1/2$. (d) Average CDW structure factor at long times, i.e., for $\tau > 10t_1^{-1}$; error bars are the standard deviation and a metric quantifying the oscillation amplitudes in the equilibrated regime.
} 
\label{fig:Fig_5}
\end{figure*} 

\section{Dynamical properties}
\label{sec:dynamics}

Having established the effects of the gain-loss non-hermiticity, i.e., a purely imaginary staggered chemical potential, in the phase diagram of the Haldane-Hubbard model, we can proceed to study its influence over real-time dynamics. For that, one should recall that non-Hermitian Hamiltonians typically arise as effective ones, valid at short times, in describing a quantum system in the presence of a bath. The Lindblad quantum master equation describes this in terms of the dynamics of the system's density matrix $\hat \rho$,
\begin{align}
    \frac{\partial\hat \rho(t)}{\partial t} = -i\left[\hat H, \hat \rho(t)\right] + \hat {\cal L}(\hat \rho)
\label{eq:master}
\end{align}
where the bath's influence in the system is mapped by $\hat {\cal L}(\hat \rho) = \gamma \sum_l \left(\hat L_l \hat \rho \hat L_l^\dagger -\frac{1}{2}\left(\hat \rho \hat L_l^\dagger\hat L_l + \hat L_l^\dagger\hat L_l\hat \rho\right)\right)$, the Liouvillian superoperator, with $\gamma>0$ setting the loss rate to the environment. If the microscopic coupling to the environment is chosen to have the particular form $\hat L_l = \sqrt{2}\hat c_l^\dagger$ ($\hat L_l = \sqrt{2}\hat c_l^{\phantom{\dagger}}$) for sites $l\in A$ ($l\in B$) sublattice, thus Eq.~\eqref{eq:master} can be rewritten, if neglecting the contribution of the term proportional to $\hat L_l \hat \rho \hat L_l^\dagger$---the quantum jump term--- as:
 \begin{equation}
    \frac{\partial\hat \rho(t)}{\partial t} \simeq -i\left[\hat H_{\rm eff} \hat \rho(t)- \hat \rho(t)\hat H_{\rm eff}^\dagger\right]\ ,
\end{equation}
where the effective Hamiltonian governing the dynamics is $\hat H_{\rm eff} \equiv \hat H - i\gamma N_s/2$, precisely the non-Hermitian Hamiltonian we have previously studied [Eq.~\eqref{eq:H_haldane_hubbard}], subtracted by a trivial constant~\footnote{Note that such a constant term is irrelevant in the description of the equilibrium properties of $\hat H_{\rm eff}$ but it is fundamental in the dynamics to guarantee that the norm of the time-evolved state decays in the course of the non-unitary evolution.}.

In general, however, such an approximation [dropping one term in Eq.~\eqref{eq:master}] is not justified at later times, and one needs to follow the dynamics of the full master equation to describe the evolution of a quantum system coupled to a bath. Yet, a relatively simple method exists for that, which instead of computing the dynamics of the density matrix, a $D^2$ object ($D$ is the Hilbert space dimension), one can tackle it via non-unitary dynamics with $\hat H_{\rm eff}$ supplemented by the stochastic inclusion of the quantum jump terms~\cite{Dalibard1992, Dum1992, Dum1992b, Plenio1998, Daley2014}. Such an approach, dubbed the quantum trajectory method, thus deals with the evolution of quantum states ($D$-sized objects), and physical quantities measured over time converge with the number of `trajectories,' i.e., a set of time-random applications of the jump operators; details of our particular implementation can be found in Ref.~\cite{Wang2023}. In what follows, physical quantities are averaged over 50 such trajectories, with the error bars estimating the temporal standard error of the mean. Additionally, we choose the initial state for the dynamics as the ground state of the hermitian case ($\gamma = 0$) and quench the gain-loss at a time $\tau = 0$, measured in units of $(t_1)^{-1}$. 
 
Figures~\ref{fig:Fig_5}(a)--\ref{fig:Fig_5}(c) illustrate the time evolution of the trajectory averaged structure factor $\overline{\langle S_{\rm cdw}\rangle}$ for different interaction strengths and contrasting various gain-loss amplitudes $\gamma$. First, the insets show that the average density hardly changes over time, oscillating about the initial density at half-filling, a direct consequence of the balanced gain-loss, unlike the typical exponential decay in time if only dissipation were considered~\cite{Wang2023}. Additionally, for $\tau \gtrsim 10t_1^{-1}$, all sets of parameters lead to an equilibration of the charge ordering, quantified by a saturation of $\overline{\langle S_{\rm cdw}\rangle}$. Remarkably, despite very different initial states, the equilibrated values for the average CDW structure factor $\overline{\langle S_{\rm cdw}\rangle}(\tau \rightarrow \infty)$ do not exhibit appreciable differences for different interaction strengths, only significant dependence on the gain-loss amplitude $\gamma$ -- see Fig.~\ref{fig:Fig_5}(d). That is, whether the system starts as a topological insulator state [$V=t_1$, Fig.~\ref{fig:Fig_5}(a)] or a Mott insulator state [$V=3t_1$, Fig.~\ref{fig:Fig_5}(c)] is irrelevant to the long-time dynamics. 

Such a monotonic increase of $\overline{\langle S_{\rm cdw}\rangle}(\tau \rightarrow \infty)$ with $\gamma$, irrespective of the $V$ values, which asymptotically approaches the saturated regime $\langle S_{\rm cdw}^{\rm max}\rangle = N_e/2= 6$ corresponding to a perfect Fock state with occupancies in only one of the sublattices, can be interpreted as the manifestation of a typical band-insulator under the dynamics. Indeed, that is what one would expect if the imaginary staggered potential, here associated with the gain-loss, was purely real. As a result, assuming that non-hermiticity is merely a construction that emerges from the coupling to a bath, the dynamics under these conditions cast doubt on whether it is physically relevant to assign regimes as CDW Mott insulators at large interaction strengths, as done in the phase diagrams of Fig.~\ref{fig:Fig_1}. This is because time-scales as small as $t_1^{-1}$ are sufficient to disrupt this order driven by $V$ [see Fig.~\ref{fig:Fig_5}(c)], and to trivially reemerge at later times being dominated by $\gamma$. The latter leads to a band-insulating phase that explicitly breaks the sublattice symmetry (even via an imaginary potential) not governed by the interaction strengths.

\section{Summary}
\label{sec:summary}
We perform an extensive characterization of the non-Hermitian Haldane-Hubbard model with gain and loss, emphasizing the exploration of its phase diagram and dynamical properties. The phase diagram reveals three distinct regimes based on different physical quantities: the topological invariant $C$, the charge density wave order parameter $S_{\rm cdw}$, and the real and imaginary parts of the energy gap $\Delta_{\rm real}$ and $\Delta_{\rm imag}$. These analyses identify several unique phases, including gapped and (real) gapless topological insulators and (trivial) Mott insulator phases. Unlike the non-interacting regime~\cite{Li2022} in which the gapless topological phase is characterized by $\Delta_{\rm real}=\Delta_{\rm imag}=0$, finite interactions result that $\Delta_{\rm imag} >0$ and $\Delta_{\rm real}=0$ while retaining the topological character quantified by a quantized Chern number.

Increasing the gain-loss magnitude can lead to two phase transitions, the first assigned as a ${\cal PT}$-symmetry breaking in the subspace of two low-lying states, driving the topological gapped to a (real) gapless phase and a second one in which a first-order phase transition in the real part of the spectrum emerges, capturing the onset of the charge-ordered Mott insulating regime. The model's dynamical properties are also investigated, focusing on the CDW structure factor's robustness over time. The findings indicate that regardless of the interaction strengths, charge ordering emerges not because of a spontaneous symmetry breaking but rather explicitly via the (imaginary) staggered potential that maps the effects of gain and loss. Lastly, we remark that emulation of Haldane models with controlled non-hermiticity is now a reality in topoelectric circuits~\cite{Jiang2024, Xie2025}, including the case where one has a staggered loss term~\cite{Jiang2024} paralleling the development of the physics here uncovered, even if in the non-interacting regime.

\begin{acknowledgments}

T.-C.Y.~acknowledges support from the Natural Science Foundation of China (Grant No.~12404285), the Zhejiang Provincial Natural Science Foundation of China (Grant No.~LQN25A040003), and the Science Foundation of Zhejiang Sci-Tech University (Grant No.~23062182-Y). R.M.~acknowledges support from the T$_c$SUH Welch Professorship Award. Numerical simulations were performed in the Tianhe-2JK at the Beijing Computational Science Research Center and with resources provided by the Research Computing Data Core at the University of Houston. This work also used TAMU ACES at Texas A\&M HPRC through allocation PHY240046 from the Advanced Cyberinfrastructure Coordination Ecosystem: Services \& Support (ACCESS) program, which is supported by U.S. National Science Foundation grants 2138259, 2138286, 2138307, 2137603, and 2138296.
\end{acknowledgments}

\vskip 1in
\appendix
\section*{Appendix: The non-interacting limit} \label{app:non_int}
\begin{figure}[htp]\centering
\includegraphics[width=1\columnwidth]{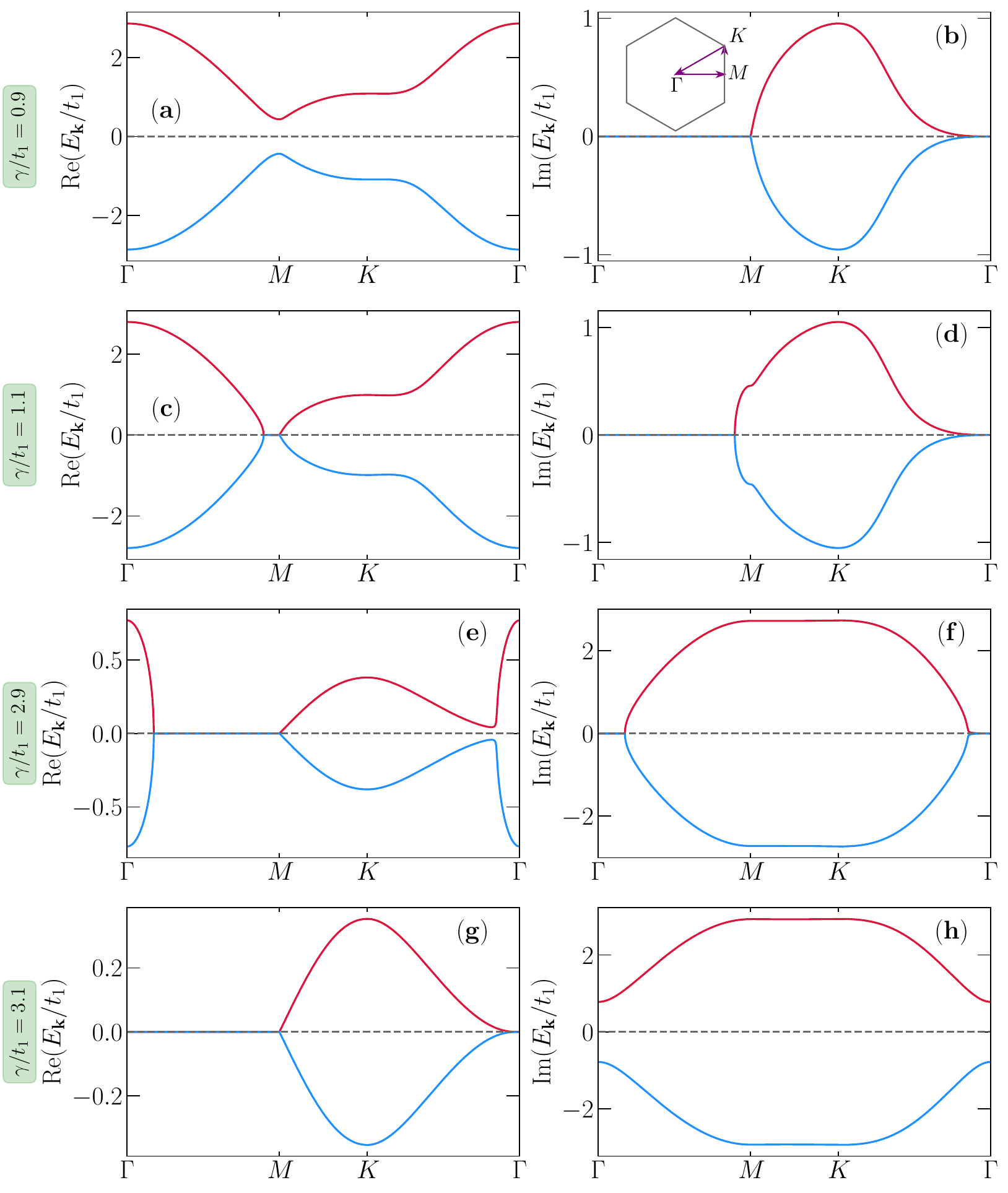}
\caption{Non-interacting band structure [Eq.~\eqref{eq:dispersion}] along a high-symmetry path in the first Brillouin Zone, depicted as an inset in (b). The left (right) panels give the real (imaginary) part of the spectrum. Panels (a--d) show the dispersion around the critical gain-loss strength that leads to the transition from the non-Hermitian topological to the topological gapless phase, $\gamma_{c1}/t_1=1$, as indicated. Panels (e--h) display the bands around the second transition, from the topological gapless phase to the non-Hermitian trivial insulating one taking place at $\gamma_{c2}/t_1=3$.}
\label{fig:bands_appendix}
\end{figure} 
The non-interacting limit ($V=0$) of the original Hamiltonian [Eq.~\ref{eq:H_haldane_hubbard}] in the presence of gain and loss reads:
\begin{align}
    \hat H_0 = &-t_1\sum_{\langle l,l^\prime\rangle}\hat c_l^\dagger \hat c_{l^\prime}^{\phantom{\dagger}} 
    -t_2\sum_{\langle\langle l,l^\prime\rangle\rangle}e^{i\phi_{ll^\prime}}\hat c_l^\dagger \hat c_{l^\prime}^{\phantom{\dagger}}\notag \\
    &+i\gamma\sum_{m\in A}\hat c_m^\dagger\hat c_m^{\phantom{\dagger}} -i\gamma\sum_{m\in B}\hat c_m^\dagger\hat c_m^{\phantom{\dagger}} ,
    \label{eq:H_haldane_gain_loss_non_interacting}
\end{align}
where the fermionic operators $\hat c_l=\hat a_l,\hat b_l$ are defined in the honeycomb lattice composed of sublattices $A$ and $B$. Taking advantage of translation invariance, one can introduce the fermionic operators in momentum space, $\hat c^\dagger_{\bf k}=\frac{1}{\sqrt{N}}\sum_{\bf k}e^{-i\bf k \cdot \bf r_l} \hat c_l^\dagger$, to represent it as $\hat H_0 = \sum_{\bf k}\psi_{\bf k}^\dagger\hat H_{\bf k}\psi_{\bf k}$, where, $\psi_{\bf k}^\dagger = (\hat a_{\bf k}^\dagger\ ,\   \hat b_{\bf k}^\dagger)$. By defining the nearest-neighbor vectors ${\bf n}_1$, ${\bf n}_2$ and ${\bf n}_3$, together with the lattice's primitive vectors ${\bf a}_1$ and ${\bf a}_2$ [see Fig.~\ref{fig:Fig_1}(a)], the following representation in momentum ensues:
\begin{align}
    \hat H_{\bf k} = 
    \begin{pmatrix}
    m_+({\bf k}) + i\gamma & f(\bf k)\\ 
    f^*({\bf k}) & m_-({\bf k}) - i\gamma
    \end{pmatrix} \ ,
\end{align}
where, $f({\bf k}) = -t_1(e^{i {\bf k} \cdot {\bf n}_1} + e^{i {\bf k} \cdot {\bf n}_2} + e^{i {\bf k} \cdot {\bf n}_3})$, and
\begin{align}
    m_{\pm}({\bf k}) = -2t_2[&\cos ({\bf k} \cdot{\bf a}_1 \pm \phi) + \cos({\bf k}\cdot{\bf a}_2 \mp \phi) \notag \\ + &\cos({\bf k} \cdot ({\bf a}_2-{\bf a}_1)\pm \phi)]\ .
\end{align}
Diagonalization results in two bands,
\begin{align}
        E_{{\bf k}} = A({\bf k}) \pm \sqrt{|f({\bf k})|^2+\Delta m({\bf k})^2 - \gamma^2 + 2i\gamma \Delta m({\bf k})} \ ,
    \label{eq:dispersion}
\end{align}
where we define $A({\bf k})\equiv [m_+({\bf k}) + m_-({\bf k})]/2$ and $\Delta m({\bf k}) \equiv [m_+({\bf k})-m_-({\bf k})]/2$. It becomes apparent that the Semenoff mass term (the second, third, and fourth terms inside the square root) exhibits both real and imaginary parts, which will thus control the real and imaginary gaps. We report in Fig.~\ref{fig:bands_appendix} cuts in the Brillouin zone, showing the evolution of the bands with the gain-loss strength close to the points at which the real part of the gap closes at $\gamma/t_1 =1$ [Fig.~\ref{fig:bands_appendix}(a,b) for $\gamma/t_1=0.9$ and Fig.~\ref{fig:bands_appendix}(c,d) for $\gamma/t_1=1.1$] and the point at which the imaginary part opens [Fig.~\ref{fig:bands_appendix}(e,f) for $\gamma/t_1=2.9$ and Fig.~\ref{fig:bands_appendix}(g,h) for $\gamma/t_1=3.1$] at $\gamma/t_1 = 3$. Compilation of the gaps across the whole Brillouin zone results in Fig.~\ref{fig:Fig_1}(b) shown in the main text.

\bibliography{refs}

\end{document}